\documentclass[12pt]{article}
\usepackage{a4wide,amsmath,amssymb,graphicx,psfrag,feynarts}

\newcommand{\gev}{\text{GeV}}
\newcommand{\tev}{\text{TeV}}
\newcommand{\fb}{\text{fb}}

\newcommand{\E}[1]{10^{#1}}

\newcommand{\MM}{\mathcal{M}}

\newcommand\Stop{\tilde t}
\newcommand\Sbot{\tilde b}

\newcommand{\mhpm}{m_{H^\pm}}

\newcommand{\mh}{m_{h^0}}
\newcommand{\mH}{m_{H^0}}
\newcommand{\mA}{m_{A^0}}
\newcommand{\MSf}{M_{\tilde f}}

\newcommand{\ee}{e^+ e^-}

\newcommand{\WH}{W^{\pm} H^{\mp}}

\newcommand{\tb}{\tan\beta}

\newcommand\ie{i.e.\ }

\newcommand\ri{\mathrm{i}}
\renewcommand\Re{\mathop{\mathrm{Re}}}

\newcommand{\MAX}{{\text{max}}}
\newcommand{\MIN}{{\text{min}}}
\newcommand{\EFF}{{\text{eff}}}

\begin{document}

\begin{titlepage}
\begin{flushright}
{IPPP/07/73\\
DCPT/07/146}
\end{flushright}
\vfill
\begin{center}
{\bf\large 
Charged Higgs-boson production in association with\\[.5cm] 
an electron and a neutrino at electron--positron colliders
}\\[2cm]
{\sc\large Oliver Brein\footnote{E-mail: Oliver.Brein@durham.ac.uk}} 
and 
{\sc \large Terrance Figy\footnote{E-mail: Terrance.Figy@durham.ac.uk}}\\[0.8cm]
{\normalsize Institute for Particle Physics Phenomenology,\\
University of Durham, DH1 3LE, Durham, United Kingdom}\\[.8cm]
\end{center}
\vfill
\begin{abstract}
We present results of a calculation of the cross section 
for the production of a charged Higgs boson in association
with an electron and a neutrino 
at electron--positron colliders
($e^+ e^- \to H^+ e^- \bar\nu_e$, $H^- e^+ \nu_e$).
We study predictions for the cross section in the 
Minimal Supersymmetric Standard Model (MSSM) and the 
Two Higgs Doublet Model (THDM), highlighting possible differences.
The process is effectively loop-induced in both models.
Hence, the cross section is expected to be strongly model-dependent.
Most notably, due to the presence of superpartners, 
the MSSM amplitude contains Feynman graphs of pentagon-type, which are 
not present in the THDM.
This is the first complete one-loop calculation of the cross section
for this process in the THDM and the MSSM. 
For both models, so far, only approximate results with limited ranges
of validity were available.
Our main aim here is to clarify several open questions 
in the existing literature on this process.
Specifically, we will discuss the validity of the Heavy Fermion loop
approximation in both models, and of the Fermion/Sfermion loop approximation
in the MSSM.
\end{abstract}
\vfill
\end{titlepage}

\section{Introduction}
The detection of a charged Higgs boson would be an unambiguous sign
of an extended Higgs sector.
With only one Higgs doublet, like in the Standard Model,
all
charged scalar degrees of freedom
present
are needed to provide the longitudinal degrees of freedom
of the charged electroweak bosons $W^\pm$.

The search for charged Higgs bosons proceeds currently at the Tevatron
and will be taken up by the LHC experiments in the near future.
For the Tevatron the only significant production mode
of charged Higgs bosons is from the decay of top quarks
in $t\bar t$ production events. 
In the presently unprobed range of charged Higgs mass values, $\mhpm$,
above the LEP limit of about $80\,\gev$ \cite{PDG07}, this is the only relevant channel
which can be reached with significant statistics given the 
luminosity and collision energy of the protons and anti-protons at the Tevatron.
The observability of this channel relies on the values of the Yukawa couplings of 
the charged Higgs boson to top and bottom quarks and to $\tau\nu_\tau$.
The former determines the production rate, while the latter enters
the branching ratios for the decays $H^\pm\to\tau\nu_\tau$.
Overproduction of $\tau$ leptons compared to the SM expectation 
is the primary signature of charged Higgs bosons.
The best $H^\pm$ search channels at hadron colliders identified so far, depend on
Yukawa interactions of heavy quarks.
Furthermore, in Two-Higgs-Doublet-Models, the $H^\pm$ Yukawa couplings depend strongly on the 
ratio of the vacuum expectation values $v_2/v_1 \equiv \tb$
of the two Higgs-doublets.
Therefore, the discovery reach for a charged Higgs boson at hadron colliders
in such models also depends strongly on the value of $\tb$. 
For instance, in the Type-II THDM and the MSSM, 
the mass reach in the intermediate $\tb$-region around $\tb=15$ is 
rather low compared to the high- and low-$\tb$ region, even at the LHC.

With an electron--positron collider, the detection of a charged 
Higgs boson is rather straightforward if its energy allows for 
$H^+ H^-$ pair production.
As the tree-level amplitude for this process consists of $s$-channel
photon and $Z$-boson exchange, the production cross section 
is essentially independent of other parameters of the model 
the charged Higgs boson is embedded in.
Furthermore, for $\mhpm$ above the top--bottom threshold, the dominant 
decay mode $H^\pm \to tb$ can be utilised to search for charged Higgs bosons
\cite{Hpair-to-tbtb},
unlike at hadron colliders where processes with these decay modes 
are swamped by background events. 
The mass-reach of the pair production process as a search channel
at an electron positron collider
is essentially independent of $\tb$ and only limited by the available 
collision energy $\sqrt s$.

It is a topic of current research to explore the mass-reach of a future 
electron positron collider for charged Higgs bosons
beyond the pair production limit of about $\sqrt s/2$.
Single production of one charged Higgs boson in association with one or a few
particles with an overall mass less than $\mhpm$ can proceed via several channels.
Unfortunately, the electroweak gauge bosons as intermediate particles 
normally do not lead to tree-level processes similar to Higgs-strahlung 
or vector-boson fusion for neutral Higgs boson production. 
The reason is that the $Z W^\pm H^\mp$ interaction 
is absent in any Higgs sector model with multiple doublets
and restricted to have a quite small coupling constant in many 
other models because of the measured electroweak $\rho$-parameter
\cite{grifols-mendez}.

The only relevant single charged Higgs boson production process
with only two particles in the final state
is the (effectively) loop-induced $W^\pm H^\mp$ production 
which has been studied in detail in the general THDM and the MSSM 
\cite{WH-papers1,WH-papers2,ownWH}.
Relevant processes with a tree-level amplitude start with three
particles in the final-state.
Among those, the final states $H^\pm t b$ \cite{tbH,singleH-KMO} 
and $H^\pm \tau \nu_\tau$ \cite{singleH-KMO,taunuH} appear to be 
the most promising.

The loop-induced process $e^+ e^- \to H^- e^+ \nu_e$ (and its charge conjugate)
have also been suggested as potential charged Higgs boson discovery processes.
Unlike in charged Higgs boson production together with $tb$ or $\tau\nu_\tau$,
this process has $t$-channel Feynman graphs, similar to the 
vector boson fusion process in the SM, which grows like $\log(s/\mhpm^2)$
for $s \gg \mhpm^2$.
The cross section for this process has been calculated in various approximations.
For the calculation of the THDM cross section
in Ref.~\cite{singleH-KMO}
only the $t$-channel Feynman graphs have been
considered with insertions of effective vertices for the 
loop-induced $\gamma W^\pm H^\mp$ and
$Z W^\pm H^\mp$ interaction. 
Among the contributions to those one-loop vertices, bosonic contributions,
i.e. contributions from virtual Higgs, Goldstone and gauge bosons, have been
checked to be small in the region where the vertex interactions are substantial
and then neglected in the numerical evaluation in \cite{singleH-KMO}.
In Ref.~\cite{eeHenu-su-etal} the THDM cross section is calculated
taking all terms quadratic in the $H^\pm t b$ Yukawa couplings into account.
This amounts to taking all one-loop Feynman graphs with
top and bottom quarks into account and neglecting all bosonic contributions.
Thus, in this study the $s$-channel contributions have been included
in the top--bottom loop approximation.
In Ref.~\cite{eeHenu-MSSM} preliminary results have been reported
on an evaluation of the 
MSSM cross section using the approximation of only taking
fermion and sfermion loops into account.
In this calculation, also all contributions from virtual gauge and Higgs bosons
have been neglected. 
Furthermore, all other contributions from virtual superpartners
which do not contain a closed sfermion loop have been neglected as well.

The current state of research concerning the process 
$e^{+}e^{-} \to H^{-} e^{+} \nu_{e}$ 
+ (c.c.) leaves quite a few open questions which we would like to address 
in this paper.
First, neglecting the bosonic contributions was justified in 
\cite{singleH-KMO,eeHenu-su-etal} by their reported smallness 
in the loop-induced $\gamma W^\pm H^\mp$ and $Z W^\pm H^\mp$ interactions
\cite{WH-papers2}.
Yet, in the process at hand there are also bosonic four-point loop graphs 
in the THDM, which have never been computed in this context.
Naturally, the MSSM also has additional four-point loops with virtual
superpartners and, furthermore, even five-point (``pentagon'') superpartner loops.
Specifically, there are triangle-, box- and pentagon-type loops
involving neutralinos, gauginos and first generation sleptons.

The rest of the paper is organized as follows.
In Section \ref{sec:process} we present the essential process
kinematics, a complete list of Feynman graphs for the process
in the MSSM and the THDM, and a description of the 
approximations which have been used so far in previous calculations
of this process.
Section \ref{sec:results} contains numerical results 
for the total cross section of $e^{+}e^{-} \to H^{-} e^{+} \nu_{e}$
+ (c.c.) 
for two interesting MSSM sample scenarios.
In this section, we demonstrate  
when certain approximations describe the complete MSSM result 
reasonably well and when (and how badly) they fail to do so.
Our conclusions follow in Section \ref{sec:conclusions}.

\section{ $e^{+}e^{-} \to H^{-} e^{+} \nu_{e}$ in the MSSM and THDM}
\label{sec:process}
\subsection{Kinematics}

We study the reaction
\begin{equation*}
  e^+(k,\lambda) + e^-(q,\bar\lambda) \to
  H^-(p) + e^+(k',\lambda') + \nu_e(q')\,,
\end{equation*} 
where 
$k$ and $q$ denote the momenta of the initial-state positron and electron,
$q'$ and $k'$ the momenta of the final-state neutrino and positron,
and $p$ the momentum of the final-state Higgs boson $H^-$.
Additionally, the electron and the
positrons, are characterized by their spin polarization 
$\lambda, \bar\lambda,  \lambda' (=\pm\frac 12)$.

The total unpolarized cross section can be written as an integral
over a 4-fold differential cross section \cite{FAFC}:
\begin{align}
\sigma & = 
\frac{1}{4}\sum_{\lambda,\bar\lambda,\lambda'=\pm 1/2}^{}
\int_{m_e}^{k'^0_\MAX} \!\!dk'^0
\int_{p^0_\MIN}^{p^0_\MAX} \!\!dp^0
\int_{\cos\theta_\MIN}^{\cos\theta_\MAX} \!\!d\cos\theta
\int_{0}^{2\pi} \!\!d\eta
\frac{d^4\sigma_{\lambda \bar\lambda \lambda'}}{dk'^0\,dp^0\,d\cos\theta\,d\eta}
\,,
\end{align}
with
\begin{align*}
k'^0_\MAX & = \frac{s-\mhpm^2+m_e^2}{2\sqrt s}
\,, & p^0_{\MAX,\MIN}&= (\sqrt s - k'^0)\frac{1+\xi}{2} \pm |\vec k'|\frac{1-\xi}{2},\\
\xi & = \frac{\mhpm^2}{s -2\sqrt s\, k'^0+m_e^2}
\,,& \cos\theta_{\MAX,\MIN}&= \pm 1 \mp \delta\cos\theta_{\text{cut}}
\,
\,.
\end{align*}
The differential cross section is related to the squared matrix element
$|\MM_{\lambda \bar\lambda \lambda'}|^2$ through
\begin{align}
\frac{d^4\sigma_{\lambda \bar\lambda \lambda'}}{dk'^0\,dp^0\,d\cos\theta\,d\eta} & =
\frac{|\MM_{\lambda \bar\lambda \lambda'}|^2}{(4\pi)^4\, s \sqrt{1-4 m_e^2/s}}
\,.
\end{align}
The angle between the three-momentum of the in- and outgoing positron, $\theta$ is 
defined by
\begin{align}
\cos\theta & = \frac{\vec k'\cdot \vec k}{|\vec k'| |\vec k|}
\,,
\end{align}
and $\eta$ is the angle between the plane spanned by the three-momenta of the 
three final-state particles and a plane perpendicular to the beam axis.

In order to have a realistic scenario concerning the detectability 
of the outgoing positron, we employ an angular cut-off 
$\delta\cos\theta_{\text{cut}} = \E{-3}$. 
This cuts off all final state positrons which scatter with an angle
less than $44.7\,$mrad to the beam axis. This choice is 
inspired by the TESLA Detector Technical Design Report 
\cite{TESLA-TDR-detector} where a detection of 
electrons and positrons down to $4.6\,$mrad is foreseen, but with angles
below $30\,$mrad only used for luminosity monitoring.

\subsection{Feynman graphs}
\label{sec:feyn}

In the MSSM and THDM 
the tree-level amplitude for the process under study
contains the $H^- e^+ \nu_e$ Yukawa coupling, which 
is $\varpropto m_{e}/m_{W} \approx 6 \cdot 10^{-6}$.
Thus, the tree-level contribution is strongly suppressed
and can be neglected.
The process can be called effectively loop-induced.
In our calculation we take into account all one-loop contributions 
to the amplitude which do not 
vanish in the limit $m_{e}=0$. For this reason, Feynman graphs 
with an insertion of an $s$-channel $Z-A$
mixing self-energy, or a $t$-channel neutrino self-energy, and 
the  radiative corrections 
to $e^{+}e^{-} \{ h^{0},H^{0},A^{0}\}$ Yukawa couplings 
and large parts to the $e^{\pm} \nu_{e} H^{\mp}$ Yukawa coupling
need not be considered. 
In the MSSM, there is a contribution to the $e^{\pm} \nu_{e} H^{\mp}$ 
Yukawa coupling which involves virtual charginos or neutralinos 
which is finite and does not vanish for $m_e = 0$.
Those contributions are taken into account in our calculation.
\smallskip

\begin{figure}[htb]
\begin{footnotesize}
\input{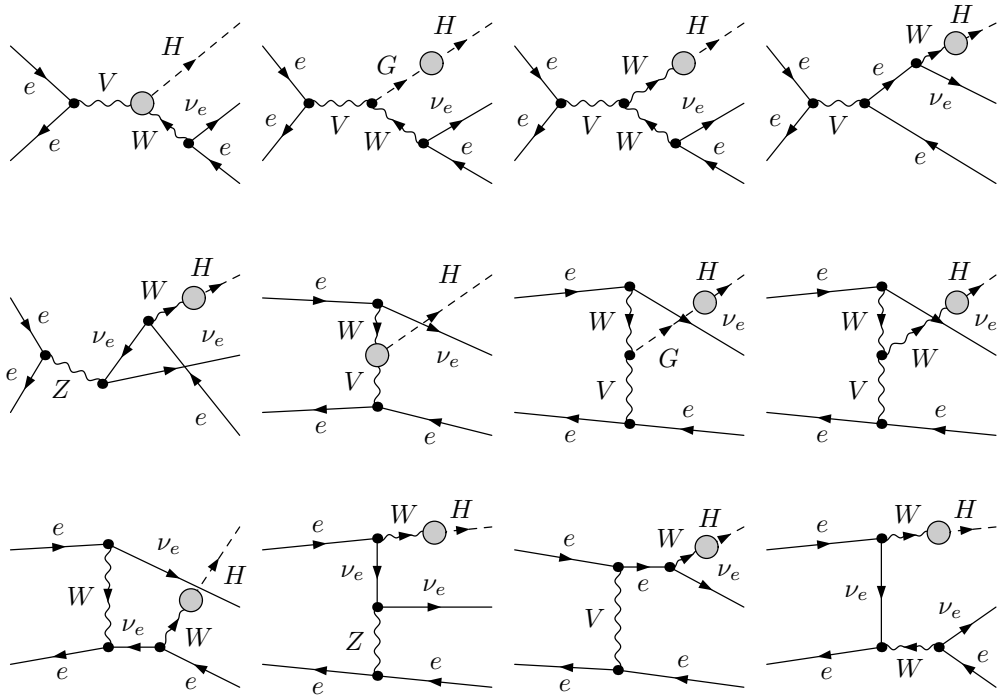}
\end{footnotesize}
\caption{
\label{fig:whself} Feynman graphs for $e^{+}e^{-} \to e^{+} \nu_{e} H^{-}$ with insertions
of self-energy and 
$\gamma W^+ H^-$ and $Z W^+ H^-$ vertex corrections. 
The counter-term graphs have exactly the 
same structure, with the loop-insertion replaced by the appropriate 
counter-term. Here $V=\gamma,Z$.
}
\end{figure}

\begin{figure}[htb]
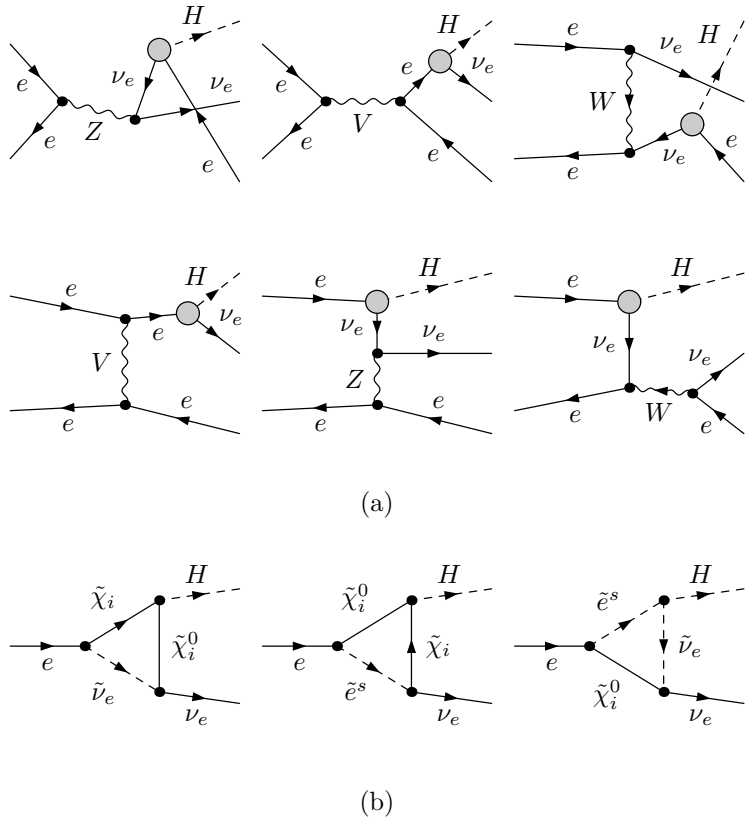

\begin{center}
\begin{footnotesize}
\input{eeHenu-yuk-vf.tex}
(a)\\
\input{enuH-vertex.tex}
(b)
\end{footnotesize}
\end{center}
\caption{
\label{fig:yuk} (a) Non-vanishing Feynman graphs for 
$e^{+}e^{-} \to e^{+} \nu_{e} H^{-}$ with 
$e^{\pm} \nu_{e} H^{\mp}$ Yukawa coupling insertions.
(b) Superpartner corrections to the $e^{\pm} \nu_{e} H^{\mp}$ Yukawa coupling
which do not vanish for $m_e = 0$. Here $V=\gamma,Z$.
}
\end{figure}

\begin{figure}[htb]
\begin{footnotesize}
\input{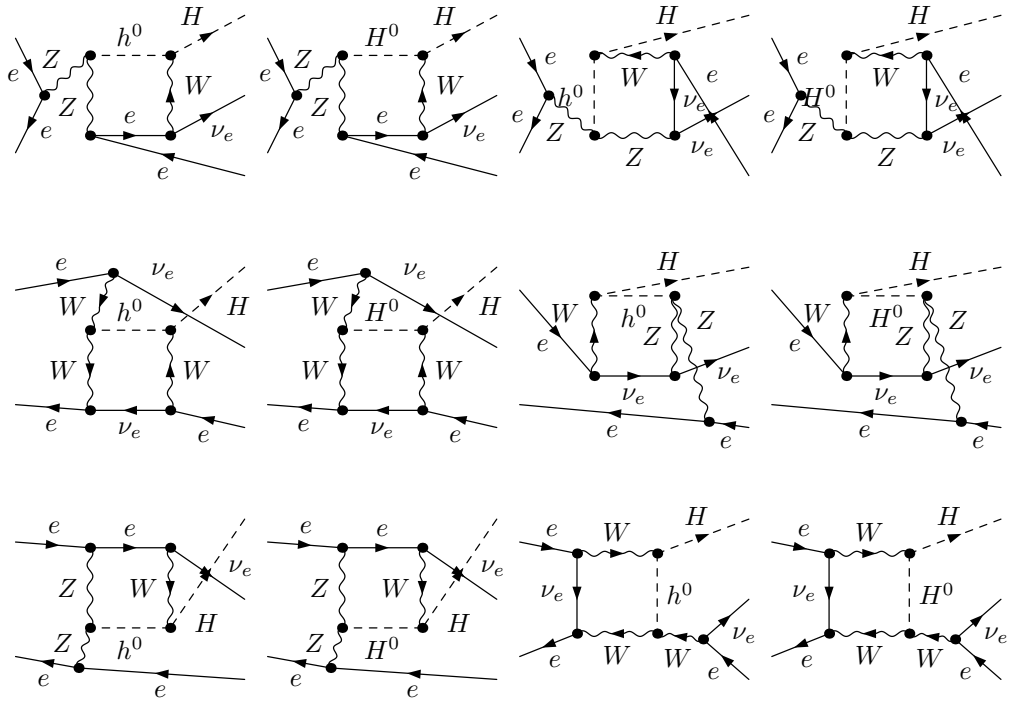}
\end{footnotesize}
\caption{\label{fig:thdmbx} Box-type Feynman graphs for  $e^{+}e^{-} \to e^{+} \nu_{e} H^{-}$ in the THDM.}
\end{figure}

\begin{figure}[htb]
\begin{footnotesize}
\input{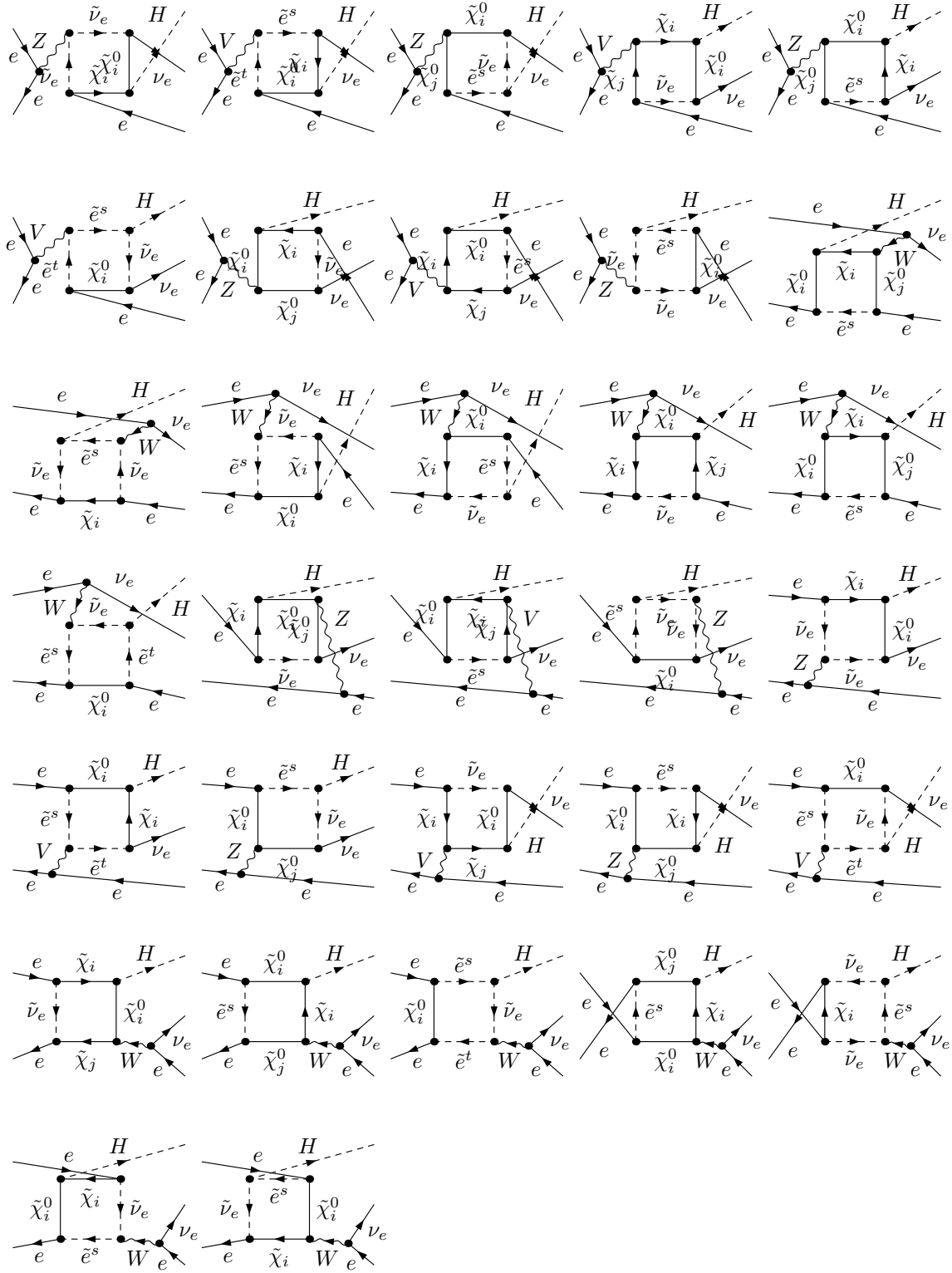}
\end{footnotesize}
\caption{\label{fig:spbox} Box-type Feynman graphs for  $e^{+}e^{-} \to e^{+} \nu_{e} H^{-}$ with 
virtual superpartners in the MSSM. Here $V=\gamma,Z$.}
\end{figure}

\begin{figure}[htb]
\begin{footnotesize}
\input{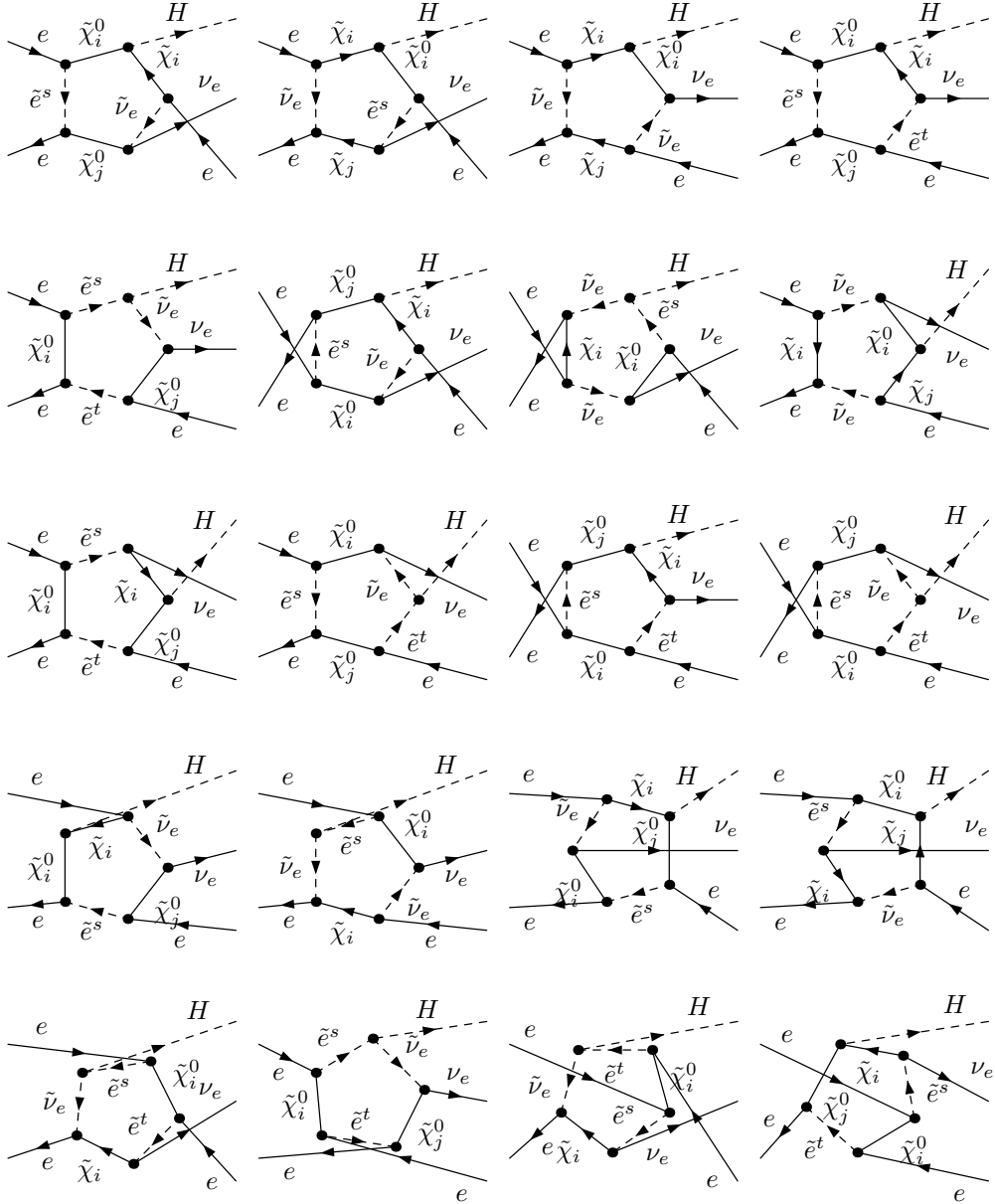}
\end{footnotesize}
\caption{\label{fig:pent} Pentagon-type Feynman graphs for  $e^{+}e^{-} \to e^{+} \nu_{e} H^{-}$ in the MSSM.}
\end{figure}

The Feynman graphs of the MSSM loop contributions can be divided into
\begin{itemize}
\item 
graphs  with $W^-$--$H^-$ and $G^-$--$H^-$ mixing self-energies on the external
$H^-$ line (see Fig.\ref{fig:whself}),
\item
graphs with the loop-induced $\gamma W^{+} H^{-}$ or $Z W^{+} H^{-}$ vertex
(see Fig.\ref{fig:whself}),
\item
graphs with the non-vanishing part of the $e^{+} \nu_{e} H^{-}$ vertex
(see Fig.\ref{fig:yuk}),
\item 
box-type graphs (see Figs.~\ref{fig:thdmbx} and \ref{fig:spbox}),
\item 
pentagon-type graphs (see Fig.\ref{fig:pent}).
\end{itemize}
The self-energy and vertex insertions depicted in Fig.\ref{fig:whself} 
consist of 
closed loops of fermions and sfermions, 
loops with gauge and Higgs bosons,
and loops with electroweak gauginos and sfermions.
\footnote{The Feynman graphs for these insertions can be found e.g.
	in \cite{ownWH}.}

In the literature there have been several approximations adopted to simplify the calculations as a complete
calculation in the MSSM requires the calculation of pentagon graphs. 
Below, we 
define the typical approximations and to what subset of Feynman graphs they
correspond to.
\begin{description}
\item[Fermion/Sfermion Approximation :] 
Only closed loops with fermions or sfermions are included in the calculation.
For the THDM part of the amplitude this approximation means 
neglecting all bosonic loops. 
The superpartner part of the amplitude comprises the set of graphs analogous
to the fermion loop amplitudes, with fermions replaced by sfermions
plus some additional related ones involving 4-point interactions with sfermions.
Such an approximation can be justified in scenarios where the masses of the 
third-generation sfermions (especially squarks) lie well below the masses of 
slepton and electroweak gaugino masses which appear in all the neglected graphs 
(see Figs.~\ref{fig:yuk}, \ref{fig:spbox} and \ref{fig:pent})
and provide additional mass suppression factors.

\item[sTHDM :] 
No superpartner graphs are included in the calculation. 
This subset of graphs corresponds to the full THDM with
the Higgs sector parameters chosen according to 
the corresponding MSSM scenario.
It is usually a useful approximation to the full MSSM result
in parameter scenarios where all superpartners are rather heavy
compared to the SM particles.

\item[Heavy Fermion Approximation :] 
Only the loops with third generation fermions
are taken into account.
This approximation consists of all THDM terms proportional to 
the usually dominant third generation Yukawa couplings and
neglects the bosonic loops of Fig.~\ref{fig:thdmbx}.
In the literature, this approximation is usually further 
simplified by only taking the loops with top and bottom quarks into account.
In the present calculation, we also include the loops involving tau leptons
which has a small but noticeable effect on the results.
\end{description}

\subsection{Calculation}

Although the tree-level contribution vanishes in the limit of
vanishing electron mass, which we consider, the need for renormalization
arises at one-loop.  
In the gauge and Higgs boson sector 
there are divergent off-diagonal propagator entries 
leading to $H^\pm$--$W^\pm$ and $H^\pm$--$G^\pm$ mixing and,
connected to this, the one-loop $\gamma W^\pm H^\mp$
and $Z W^\pm H^\mp$ vertex functions are divergent too 
(see Fig.~\ref{fig:whself}).
We use the on-shell renormalization scheme of
Ref.~\cite{onshell-SUSY-Higgs}, the application of which to
the present situation is discussed e.g. 
in~\cite{WH-papers1,WH-papers2,ownWH}.  
In this scheme the renormalization conditions relevant to 
our calculation are the following.
\begin{itemize}
\item
Renormalized tadpole graphs for the neutral CP-even Higgs bosons,
$h^0, H^0$, vanish:
\begin{align}
  \label{eq:tadpole-cond1}
  {\hat t}_{h^0} &= t_{h^0} + \delta t_{h^0} = 0\,, \\
  \label{eq:tadpole-cond2}
  {\hat t}_{H^0} &= t_{H^0} + \delta t_{H^0} = 0\,.
\end{align}
This guarantees that the parameters $v_1$, $v_2$ in the renormalized
Lagrangian describe the minimum of the Higgs potential at one-loop 
order.

\item
Physical charged Higgs bosons $H^\pm$ do not mix with longitudinally 
polarized $W^\pm$ bosons, \ie the real part of the renormalized 
$H^\pm$--$W^\mp$ mixing self-energy\footnote{
	The renormalized $H^\pm$--$W^\pm$ mixing self-energy is defined
	as the coefficient of $-\ri\frac{k^\mu}{m_W}$ of the amputated
	renormalized $H^\pm$--$W^\pm$ propagator.},
\begin{equation}
  \hat\Sigma_{HW}(k^2) = \Sigma_{HW}(k^2) - m_W^2 \delta Z_{HW}\,,
\end{equation}
vanishes if the momentum $k$ of $H^\pm$ is on mass-shell:
\begin{equation}
  \label{eq:HW-cond}
  \Re\hat\Sigma_{HW}(k^2) \big|_{k^2 = \mhpm^2} = 0\,.
\end{equation}
\end{itemize}
Conditions (\ref{eq:tadpole-cond1}) and (\ref{eq:tadpole-cond2})
amount to neglecting all Feynman graphs with tadpoles in our calculation.
The specific expressions for the tadpole counter-terms are not needed here.
Condition (\ref{eq:HW-cond}) fixes the renormalization constant 
$\delta Z_{HW}$:
\begin{equation}
  \delta Z_{HW} = \frac 1{m_W^2} \Re\Sigma_{HW}(\mhpm^2)\,.
\end{equation}
Furthermore, the renormalization of the divergent $H^\pm$--$G^\pm$ 
mixing self-energy is connected to the $H^\pm$--$W^\pm$ mixing self-energy
through a Slavnov--Taylor identity \cite{CGGJS}:
\begin{equation}
  k^2 \hat\Sigma_{HW} (k^2) - m_W^2 \hat\Sigma_{HG}(k^2) = 0 \;.
\end{equation}
As a consequence, the real part of the renormalized $H^\pm$--$G^\pm$ 
mixing self-energy,
\begin{equation}
  \hat\Sigma_{HG}(k^2) = \Sigma_{HG}(k^2) - k^2 \delta Z_{HG}\,,
\end{equation}
also vanishes for $k^2 = \mhpm^2$:
\begin{equation}
  \Re\hat\Sigma_{HG}(k^2) \big|_{k^2 = \mhpm^2} = 0
\,,
\end{equation}
and fixes the renormalization constant $\delta Z_{HG}$:
\begin{equation}
\delta Z_{HG} = -\Re\Sigma_{HG}(\mhpm^2)/\mhpm^2
\,.
\end{equation}
The Feynman rules for the corresponding counter-term interactions read:
\begin{align}
\label{eq:ct-wh}
\Gamma_{\text{CT}}[H^\mp W^\pm(k^\mu)] &= 
        \ri \frac{k^\mu}{m_W} m_W^2\, \delta Z_{HW}
\,,\\
\label{eq:ct-gwh}
\Gamma_{\text{CT}}[\gamma_\mu W^\pm_\nu H^\mp] & = 
        -\ri e m_W g_{\mu\nu}\,\delta Z_{HW}
\,,\\
\label{eq:ct-zwh}
\Gamma_{\text{CT}}[Z_\mu W^\pm_\nu H^\mp] &=
          \ri e m_W \frac{s_w}{c_w} g_{\mu\nu}\, \delta Z_{HW}
\,,\\
\label{eq:ct-gh}
  \Gamma_{\text{CT}}[H^\mp G^\pm (k) ] &= 
	\ri k^2 \delta Z_{HG}
\,,
\end{align}
where $Z, W^\pm$ and $\gamma$ denote the electroweak gauge bosons
and the photon, and $k^\mu$ the momentum of the $W^\pm$ boson, chosen
as incoming.

The calculation of the amplitude has been performed using 
the 't~Hooft--Feynman
gauge and Constrained Differential Renormalization \cite{CDR} with the
help of the computer programs FeynArts 3.2 and FormCalc 5.1 \cite{FAFC}.  
In particular, we made use of the numerical routines
for the evaluation of the 5-point loop-integrals implemented in
LoopTools 2.2 \cite{looptools} employing the methods of \cite{denner-dittmaier-5p}.
For our purpose, we extended the existing FeynArts model file for the MSSM
by including the necessary counter-term definitions and Feynman rules for the 
counter-term interactions.

A subclass of Feynman graphs corresponds to the production of 
a charged Higgs boson $H^-$ in association with a virtual $W^+$
and its subsequent decay into $e^+ \nu_e$. Those graphs can become 
resonant which requires to take the width of the $W$-boson into account.
We have chosen the finite width scheme and have introduced a 
constant finite width $\Gamma_W$
by replacing in the 't Hooft-Feynman gauge:
\begin{align*}
\frac{-ig^{\mu\nu}}{p^2 -m_W^2} \to \frac{-ig^{\mu\nu}}{p^2 -m_W^2 +i m_W \Gamma_W}
\,.
\end{align*}

\section{Results}
\label{sec:results}

\subsection{Parameter scenarios}
\label{parameter-scenarios}

We pick the MSSM parameter scenarios from \cite{ownWH} 
which are partly modifications
from the LEP Higgs search benchmark scenarios \cite{LEP-susy-Higgs}: 
the $m_h^\MAX$ scenario with a lower sfermion mass scale 
and the small-$\alpha_\EFF$ scenario.
This will allow us to re-use parts of the discussion of the behaviour 
of $\sigma(e^+ e^-\to \WH)$ in those scenarios from \cite{ownWH}.
The two MSSM parameter scenarios are specified as follows:
\begin{description}
\item{{\em $m_h^\MAX (400)$ scenario:}}
The soft-breaking sfermion mass parameter is
set to $\MSf = 400\,\gev$.
The off-diagonal term $X_t$ ($= A_t-\mu\cot\beta$)
in the top-squark mass matrix is set to $2 \MSf$ ($=800\,\gev$)
The Higgsino and gaugino mass parameters have the settings
$\mu=- 200 \,\gev$,
$M_1 = M_2 = 200 \,\gev$,
$M_{\tilde g} = 800 \,\gev$.
When $\tb$ is changed, $A_t$ is changed accordingly to ensure $X_t = 2 \MSf$. 
The settings of the other soft-breaking scalar-quark Higgs couplings 
are $A_b = A_t$ and $A_f = 0$ ($f=e,\mu,\tau,u,d,c,s$).

\item{{\em small-$\alpha_\EFF$ scenario:}}
This scenario gives rise to suppressed branching ratios for the
decays $h^0\to b\bar b$
and $\tau^+ \tau^-$, especially
for large $\tb$ and moderate values of $m_A$.
The settings are: $\MSf=800\,\gev$,
$X_t=-1100\,\gev$,
$M_1=M_2=500\,\gev$, $\mu=2000\,\gev$.
Also here, $A_t$ is changed if $\tb$ changes in order to keep the
value of $X_t$ fixed, $A_b=A_t$ and $A_f = 0$ ($f=e,\mu,\tau,u,d,c,s$).
\end{description}
The resulting masses of the Higgs bosons and the relevant superpartner
particles for these parameter choices 
are given in Table~\ref{tab:higgs-n-SP-masses}.

\begin{table}[t]
\begin{center}
\begin{tabular}{r||r|r|r|r}
 & \multicolumn{2}{|c|}{$m_h^\MAX (400)$ scenario} 
	& \multicolumn{2}{|c}{small-$\alpha_\EFF$ scenario} \\
 & $\tb=5$ & $\tb=30$ & $\tb=5$ & $\tb=30$\\
\hline
\multicolumn{5}{l}{Higgs masses for $\mhpm=250\,[500]\,\gev$}\\
\hline
$\mh\; [\gev]$ & 115.5 [116.9] & 122.1 [122.0]
	& 113.6 [113.9] & 119.9 [119.2]\\
$\mH\; [\gev]$ &  240.7 [494.7] &  232.3 [491.3]
        &  237.2 [493.1] & 234.0 [492.1]\\
$\mA\; [\gev]$ &  237.1 [493.5] &  232.6 [492.2]
        & 238.9 [494.4] & 238.3 [494.4]\\
\hline
\multicolumn{5}{l}{Stau, sbottom and stop masses}\\
\hline
$m_{\tilde\tau_1}\; [\gev]$ & 398.5 & 387.3 & 789.2 & 730.5 \\
$m_{\tilde\tau_2}\; [\gev]$ & 406.3 & 417.3 & 813.0 & 866.3 \\
\hline
$m_{\Sbot_1}\; [\gev]$ & 391.9 & 360.7 & 769.2 & 596.0 \\
$m_{\Sbot_2}\; [\gev]$ & 412.7 & 440.5 & 832.0 & 963.8 \\
\hline
$m_{\Stop_1}\; [\gev]$ & 224.4 & 224.0 & 692.0 & 691.9 \\
$m_{\Stop_2}\; [\gev]$ & 569.7 & 569.6 & 925.2 & 925.1 \\
\hline
\multicolumn{5}{l}{Chargino and neutralino masses}\\
\hline
$m_{\chi^\pm_1}\; [\gev]$ & 166.3 & 153.5 &  497.8 &  498.9\\
$m_{\chi^\pm_2}\; [\gev]$ & 255.5 & 263.4 & 2003.8 & 2003.5\\
\hline
$m_{\chi^0_1}\; [\gev]$   &  93.9 &  89.8 &  236.8 &  237.1\\
$m_{\chi^0_2}\; [\gev]$   & 163.8 & 154.9 &  497.8 &  498.9\\
$m_{\chi^0_3}\; [\gev]$   & 215.1 & 211.9 & 2001.1 & 2001.6\\
$m_{\chi^0_4}\; [\gev]$   & 252.4 & 262.1 & 2003.7 & 2002.9\\
\hline
\end{tabular}
\end{center}
\caption{\label{tab:higgs-n-SP-masses}
Masses of the Higgs bosons, the most $\tb$-sensitive sfermions, 
charginos and neutralinos
for the two sample scenarios and different values of $\tb$ and $\mhpm$.
All other sfermion masses are equal to $\MSf$ within $\pm 2\%$.}
\end{table}

\subsection{Cross sections}

\begin{figure}[hbt]
\centerline{\includegraphics{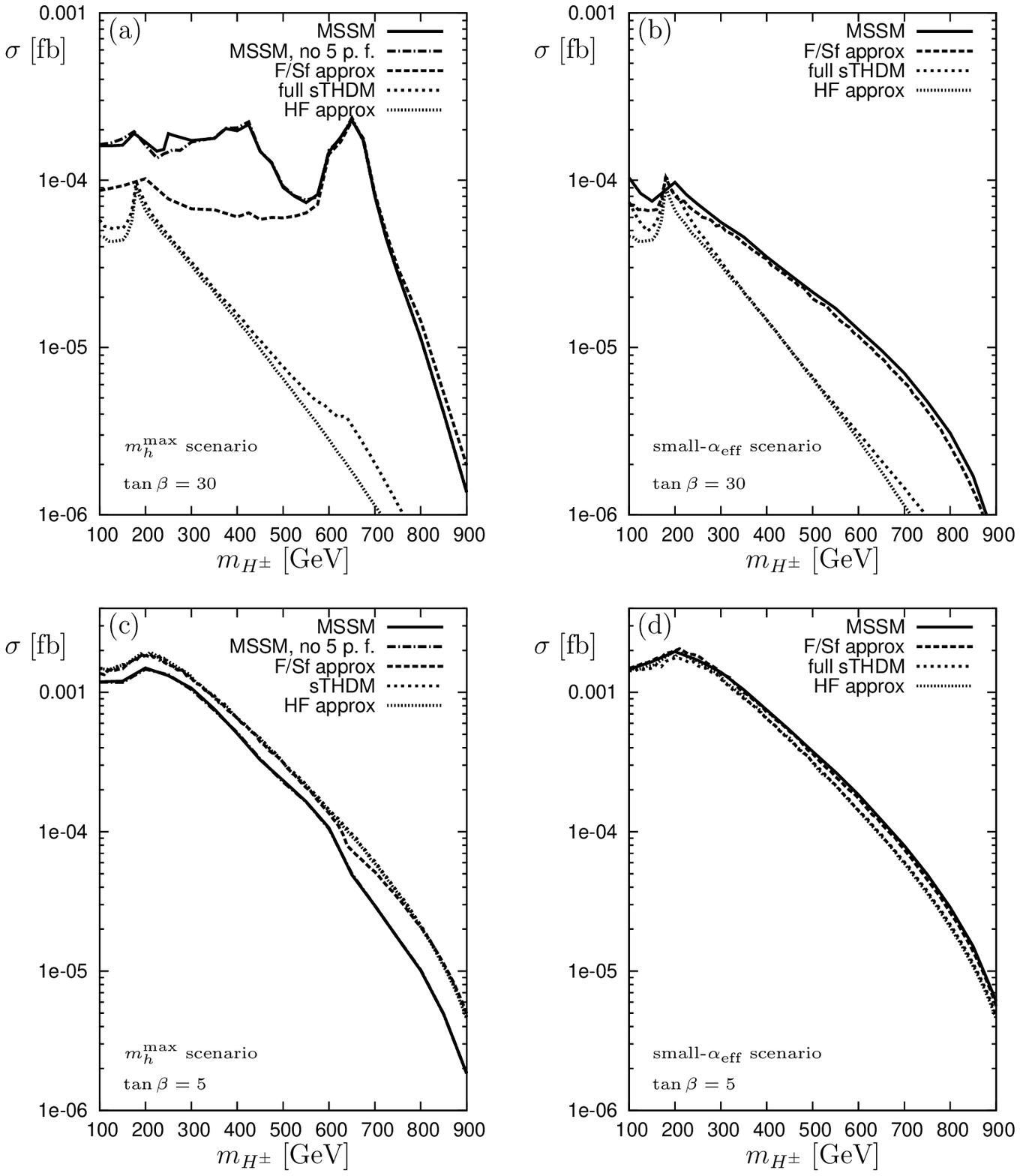}}
    \caption{
	\label{fig:mhpm-plot}
	Cross section for the process 
        $\ee \to (H^+ e^- \bar\nu_e, H^- e^+ \nu_e)$ as a function
	of a charged Higgs mass $\mhpm$ for the $m_h^\MAX (400)$ scenario 
	and the small-$\alpha_\EFF$ scenario 
	for $\tb=30$ and 5.
	Curves are shown for the full MSSM (solid lines), 
	the Fermion/Sfermion approximation (long dashed lines), 
	the full sTHDM (dashed lines), 
	and the Heavy Fermion approximation (dotted lines).  For the
        $m_{h}^{max}(400)$ scenario we also show the MSSM result
	without pentagons (dot-dashed lines).
        }
\end{figure}

\begin{figure}[hbt]
\centerline{\includegraphics{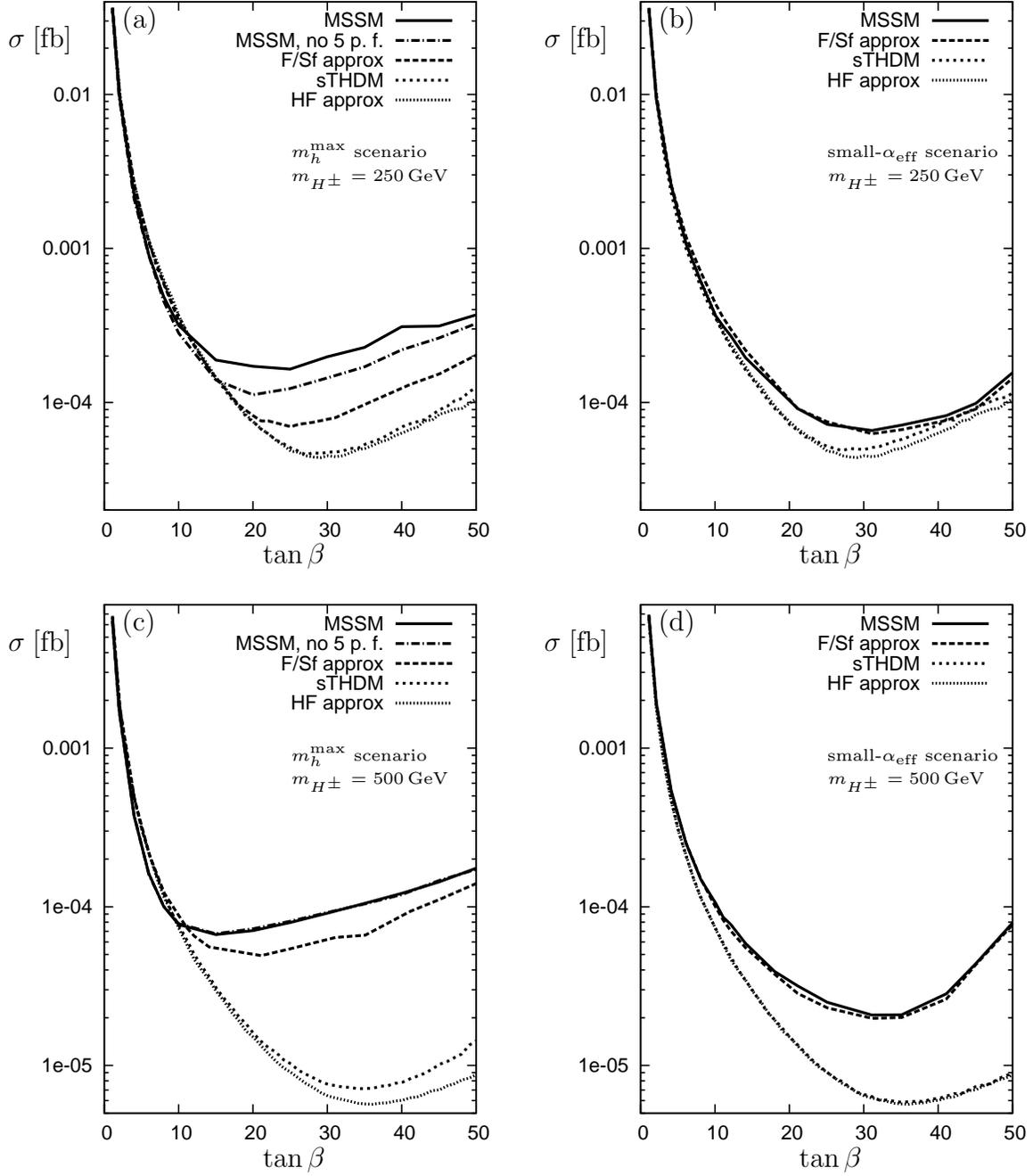}}
    \caption{
	\label{fig:tb-plot}
	Cross section for the process 
        $\ee \to (H^+ e^- \bar\nu_e, H^- e^+ \nu_e)$ as a function
	of $\tb$ for the $m_h^\MAX (400)$ scenario 
	and the small-$\alpha_\EFF$ scenario 
	for $\mhpm=250\,\gev$ and $500\,\gev$.
	Curves are shown for the full MSSM (solid lines), 
	the Fermion/Sfermion approximation (long dashed lines), 
	the full sTHDM (dashed lines), 
	and the Heavy Fermion approximation (dotted lines).  For the
        $m_{h}^{max}(400)$ scenario we also show the MSSM result
        without pentagons (dot-dashed lines).
        }
\end{figure}

We present results for 
the cross section of the process $e^+ e^- \to (H^+ e^- \bar\nu_e, H^- e^+ \nu_e)$
assuming unpolarized beams\footnote{
From the analysis of the related process $e^+ e^- \to H^\pm W^\mp$ \cite{ownWH},
we expect that the cross section can be up to a factor of 4 higher than
in the unpolarized case, if optimal polarization of the $e^-$ and
$e^+$ beams is assumed.} and 
a collider energy of $1\,\tev$. 
Assuming the collider would accumulate 1000 events/fb of integrated luminosity,
a cross section of $0.05\,\fb$ would result in 50 expected events.
This can be taken as a reasonable lower limit for the observability
of a particular discovery channel and is often not reached in our sample scenarios.
However, our aim here is not primarily phenomenology but to clarify
the relationship between different commonly used approximations 
and the complete MSSM result.
Specifically, we show cross section results for 
the $m_h^\MAX (400)$ scenario 
(Figs.~\ref{fig:mhpm-plot}a, \ref{fig:mhpm-plot}c, 
 \ref{fig:tb-plot}a, \ref{fig:tb-plot}c) and 
the small-$\alpha_\EFF$ scenario 
(Figs.~\ref{fig:mhpm-plot}b, \ref{fig:mhpm-plot}d, 
 \ref{fig:tb-plot}b, \ref{fig:tb-plot}d)
and demonstrate when certain approximations describe the complete result 
reasonably well and when (and how badly) they fail to do so.

\subsubsection{$m_h^\MAX (400)$ scenario}

The $m_h^\MAX (400)$ scenario has rather light squarks, neutralinos and charginos
compared to the assumed collider energy of 1$\,\tev$.
Therefore, all these particles can contribute significantly to
the amplitude of the process. 
Especially, there are enhanced couplings of third-generation squarks 
to the charged Higgs bosons.

In Fig.~\ref{fig:mhpm-plot}a, we show the integrated cross section as a
function of charged Higgs mass, $m_{H^{\pm}}$, for $\tan \beta=30$.
Depending on $\mhpm$, the cross section prediction of the full MSSM (solid line)
can be up to two orders of magnitude larger than of the sTHDM (dotted line).
For rather large $\mhpm$, between 550$\,\gev$ and 750$\,\gev$,
the cross section experiences threshold enhancement basically through
stop--sbottom loop graphs, which dominate the amplitude in this region,
as has been noted for the related process $e^+ e^-\to W^\pm H^\mp$
in \cite{ownWH}.
Naturally, in this region, the Fermion/Sfermion approximation used 
in \cite{eeHenu-MSSM} comes quite close to the full MSSM prediction.
For $\mhpm$ below 550$\,\gev$ this approximation ceases to be a 
good approximation to the full MSSM.
Away from the region where the loop graphs with stops and sbottoms dominate,
the ones with neutralinos and charginos become relevant as well
and lead to a cross section more than twice as large as 
in the Fermion/Sfermion approximation.

The Heavy Fermion approximation works well as an approximation 
to the sTHDM for $\mhpm$ above the threshold $\mhpm = m_t +m_b$
in top--bottom loops up to about 500$\,\gev$ where the full sTHDM
amplitude starts to experience enhancement from box-type Feynman graphs which 
consist of loops with virtual gauge and Higgs bosons (see Fig~\ref{fig:thdmbx}).
Around the peak of this effect, at $\mhpm\approx 640\,\gev$, 
the full sTHDM result lies almost a factor of two 
above the Heavy Fermion approximation.
But also for low $\mhpm$ ($\lesssim 150\,\gev$), the result in the full sTHDM 
is about 15\% larger than in the Heavy Fermion approximation.

In Fig.~\ref{fig:mhpm-plot}c, we show the integrated cross section as a
function of charged Higgs mass, $m_{H^{\pm}}$, for $\tan \beta=5$.
For small $\tb$ the difference between the full MSSM and the Fermion/Sfermion 
approximation becomes less dramatic than in the high-$\tb$ case.
However, there is, for all displayed $\mhpm$, a clear difference between the 
two results, the full MSSM lying between $-13\%$ to $-50\%$ below the 
approximation. 
The sTHDM result agrees perfectly with the Heavy Fermion approximation 
and both agree quite good with the Fermion/Sfermion approximation,
except for $\mhpm$ between 600 and 700$\,\gev$,
which is the region of stop--sbottom thresholds.
Unlike in the high $\tb$ case, there is no peak but a dip, which is also 
quite similar to the findings for $e^+ e^- \to H^\pm W^\mp$ \cite{ownWH}.

The Figures~\ref{fig:tb-plot}a and \ref{fig:tb-plot}c show the 
$\tb$ dependence of the cross section for $\mhpm=250\,\gev$ and 
$500\,\gev$ respectively.
The case $\mhpm=250\,\gev$ is particularly interesting, as it
lies exactly in the area of Fig.~\ref{fig:mhpm-plot}a 
where even the pentagon-type graphs make a noticeable 
contribution to the amplitude.
We see again that the Fermion/Sfermion approximation does not 
describe the full MSSM well. For $\mhpm=250\; [500]\,\gev$ it deviates by 
about $+30\%\;[40\%]$ for small $\tb$ and $-50\%\; [-23\%]$ for large $\tb$.
The Heavy Fermion approximation comes quite close to the 
full sTHDM result, except for the large $\tb$ region for $\mhpm=500\,\gev$
which is where the box amplitudes become enhanced as noted above.
However, there is no place in the displayed scenarios 
where the Heavy Fermion approximation is 
a reasonable approximation to the full MSSM, except for very small $\tb$
(say $\tb < 2$) and, accidentally, at a cross-over point around $\tb=10$.

\subsubsection{small-$\alpha_\EFF$ scenario}

In the small-$\alpha_\EFF$ scenario, all the sfermion thresholds in
the amplitude lie well above 1$\,\tev$ and the lowest chargino--neutralino
threshold lies above 700$\,\gev$. Hence, the superpartner effects in this 
scenario are a bit milder than in the previous one.
However, once the mass scale $\mhpm$ approaches the level of the sfermion mass scales,
stop--sbottom loop graphs can still lead to large effects if $\tb$ is large.

In Figures~\ref{fig:mhpm-plot}b and \ref{fig:mhpm-plot}d
we show the integrated cross section as a
function of $m_{H^{\pm}}$ for $\tan \beta=30$ and 5, respectively.
For large $\tb$ (see Fig. ~\ref{fig:mhpm-plot}b) there 
is no point where the Heavy Fermion approximation is a good 
approximation to the full MSSM result. 
This can only be possible for scenarios with a 
superpartner mass scale considerably higher than 800$\,\gev$.
However, for the most part 
the Fermion/Sfermion approximation works reasonably well to describe the full MSSM 
and 
the Heavy Fermion approximation likewise to describe the full THDM.
An exception is the low charged Higgs mass region around
$m_{H^{\pm}} \lesssim 300~{\rm GeV}$ where the gauge and Higgs boson
loops, which are neglected in both approximations (i.e. also in all
previous calculations of this process), contribute significantly, raising
the full MSSM and sTHDM result by up to $\approx 50\%$ compared 
to their approximations.
For small $\tb$ (see Fig. ~\ref{fig:mhpm-plot}d) the Fermion/Sfermion approximation
reproduces the full MSSM result over the whole displayed range 
within at most $\pm 10\%$ and the agreement between the Heavy Fermion approximation
and the sTHDM is similar. 
Yet, the MSSM result is usually larger than the sTHDM result, up to about $30\%$.

The $\tb$ dependence of the cross section, shown 
in Figs.~\ref{fig:tb-plot}b and \ref{fig:tb-plot}d for 
$\mhpm=250\,\gev$ and $500\,\gev$ respectively,
merely underline the statements above.

\subsubsection{The pentagon contribution}

So far, not many MSSM processes which include pentagon-type 
Feynman graphs of superpartners have been studied. 
Therefore, a closer look at this part of the amplitude seems to
be on order. 
Although the main contribution of the pentagon graphs to the 
cross section of the full MSSM comes from the interference of those 
graphs with the remainder of the MSSM amplitude, we will study 
here the cross section obtained from squaring the pentagon graphs
alone for clarity. 

\begin{figure}[hbt]
\centerline{\includegraphics{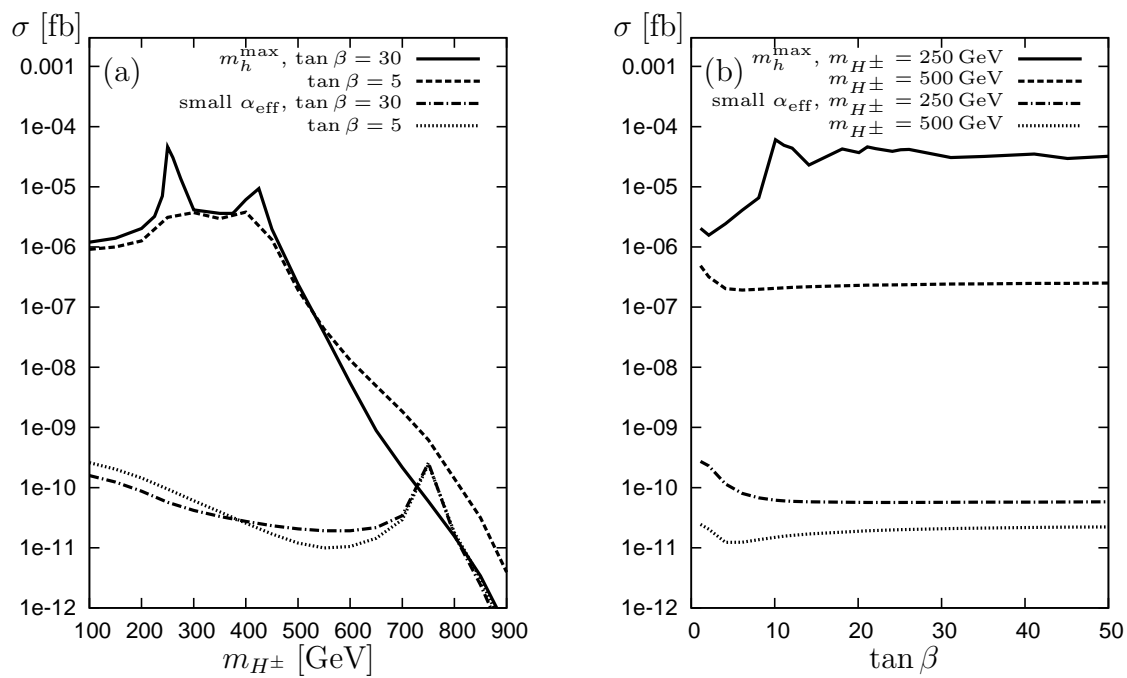}}
    \caption{
	\label{fig:pent-plot}
	Cross section value for the process 
        $\ee \to (H^+ e^- \bar\nu_e, H^- e^+ \nu_e)$ 
	resulting from 
	only using the pentagon graphs in the amplitude
	as a function of (a) $\mhpm$ and (b) $\tb$.
	Shown are results for the $m_h^\MAX (400)$ scenario
	(solid and dashed lines)
	and the small-$\alpha_\EFF$ scenario
	(dot-dashed and dotted).
        }
\end{figure}

In Fig.~\ref{fig:pent-plot}a 
we show 
this cross section as a function of $\mhpm$
for all the parameter scenarios we studied above.
For $\tb=30$ in the $m_h^\MAX (400)$ scenario we see 
threshold peaks for $\mhpm \approx m_{\chi^\pm_i}+m_{\chi^0_j}$
for some but not all combinations of $i$ and $j$, 
depending on the couplings and mixing matrices of the charginos and
neutralinos. 
For $\mhpm$ in the vicinity of the lowest 
threshold, $m_{\chi^\pm_1}+m_{\chi^0_1} = 243.3\,\gev$,
one even gets a noticeable contribution to the full result.
In the small-$\alpha_\EFF$ scenario the same threshold lies 
around $740\,\gev$ (see Fig.~\ref{fig:pent-plot}a), but in this case
the pentagon contribution is way to small to influence 
the numerical value of the cross section significantly.

The $\tb$ dependence of the ``pentagon-only'' cross section is 
displayed in Fig.~\ref{fig:pent-plot}b. 
The case of $\mhpm = 250\,\gev$ in the $m_h^\MAX (400)$ scenario
shows the most interesting behaviour. 
The cross section rises by more than an order 
of magnitude with $\tb$ going through a peak around $\tb=10$.
The reason for this is, that the neutralino and chargino masses 
depend slightly on $\tb$ and the sum $m_{\chi^\pm_1}+m_{\chi^0_1}$
passes through the value $250\,\gev$ around $\tb=10$.
The value of $m_{\chi^\pm_1}+m_{\chi^0_1}$ changes quickly for small
$\tb$ and reaches a plateau value close to $250\,\gev$ for high
$\tb$. This is why the cross section reaches a plateau for high $\tb$ 
as well.

In our sample scenarios, the MSSM is usually well approximated 
by neglecting the pentagon graphs, except for the threshold 
enhancement situations mentioned above. 
In MSSM scenarios with higher masses for the charginos, neutralinos
and sleptons than in our sample scenarios, 
with
no thresholds within the reach of the collider,
this statement should hold without exception.

Neglecting the pentagon graphs in such scenarios
is then also quite time saving from a computational point of view. 
The inclusion of the pentagons graphs adds 1216 distinct 5-point integrals 
to the amplitude to be calculated and the evaluation of those 
uses the reduction of each 5-point integral to five 4-point integrals.
This has to be contrasted to only 52 distinct 4-point integrals 
to be calculated for the amplitude without pentagons.

\section{Conclusions}
\label{sec:conclusions}

We have presented the first complete one-loop calculation 
of the process $e^{+}e^{-} \to H^{-}e^{+}\nu_{e}$ + (c.c.)
in both the THDM and MSSM. 
We have examined the differences that arise in using 
the various approximations which have been used in calculations
of this process previously
as outlined in Section~\ref{sec:feyn}.  
We find that in the small-$\alpha_{\EFF}$ scenario, that cross sections computed 
in the Fermion/Sfermion approximation are comparable to those computed in the 
full MSSM while cross sections computed in the Heavy Fermion approximation 
are comparable to those computed in the full THDM.  
However, these approximations simply fail for the $m_{h}^{max}(400)$ scenario. 
The reason is that in this scenario the virtual particles in the loops have 
lower masses which is why their contribution to the amplitude cannot be neglected. 
We further point out that cross sections computed in the Heavy Fermion 
approximation for both scenarios usually disagree completely with 
cross sections computed in the MSSM, except for very low $\tb$. 

For our sample scenarios we explicitly show
that the pentagon graphs in the MSSM 
usually contribute negligibly to the total cross section.
The MSSM is well approximated in the small-$\alpha_{\EFF}$ scenario
by neglecting the pentagon graphs.
This is also true for the $m_{h}^{max}(400)$ scenario except
in threshold regions.

\section*{Acknowledgements}

Work supported in part by the European Community's Marie-Curie Research
Training Network under contract MRTN-CT-2006-035505
`Tools and Precision Calculations for Physics Discoveries at Colliders'.

\begin{flushleft}

\end{flushleft}

\end{document}